\documentclass[11pt]{article}
\usepackage{amsmath,amssymb,color,graphics,epsfig}
\usepackage{graphicx}
\usepackage{subfigure}

\usepackage{amsfonts}

\newcommand{\be}{\begin{equation}}
\newcommand{\ee}{\end{equation}}
\newcommand{\bea}{\setlength\arraycolsep{2pt} \begin{eqnarray}}
\newcommand{\eea}{\end{eqnarray}}
\newcommand{\nn}{\nonumber}
\newcommand{\mc}{\mathcal}
\newcommand{\mm}{\mathrm}

\def\ft#1#2{{\textstyle{\frac{\scriptstyle #1}{\scriptstyle #2} } }}
\def\fft#1#2{{\frac{#1}{#2}}}

\def\0{{\sst{(0)}}}
\def\1{{\sst{(1)}}}
\def\2{{\sst{(2)}}}
\def\3{{\sst{(3)}}}
\def\4{{\sst{(4)}}}
\def\5{{\sst{(5)}}}
\def\6{{\sst{(6)}}}
\def\7{{\sst{(7)}}}
\def\8{{\sst{(8)}}}
\def\sst#1{{\scriptscriptstyle #1}}

\def\pa{{\partial}}

\begin{document}

\begin{flushright}
\end{flushright}

\vspace{25pt}
\begin{center}
{\large {\bf  Momentum-Krylov complexity correspondence} }

\vspace{10pt}
 Zhong-Ying Fan$^{1\dagger}$\\

\vspace{10pt}
$^{1\dagger}${ Department of Astrophysics, School of Physics and Material Science, \\
 Guangzhou University, Guangzhou 510006, China }\\


\vspace{40pt}

\underline{ABSTRACT}
\end{center}

In this work, we relate the growth rate of Krylov complexity in the boundary to the radial momentum of an infalling particle in AdS geometry. We show that in general AdS black hole background, our proposal captures the universal behaviors of Krylov complexity at both initial and late times. Hence it can be generally considered as an approximate dual of the Krylov complexity at least in diverse dimensions. Remarkably, for BTZ black holes, our holographic Krylov complexity perfectly matches with that of CFT$_2$ at finite temperatures.

\vfill {\footnotesize  Email: $^\dagger$fanzhy@gzhu.edu.cn\,,}

\thispagestyle{empty}

\pagebreak

\tableofcontents
\addtocontents{toc}{\protect\setcounter{tocdepth}{2}}




\section{Introduction}

  The Krylov complexity ( K-complexity) was first introduced in \cite{Parker:2018yvk} to describe the Heisenberg evolution of operators $\mc{O}(t)=e^{iHt}\mc{O} e^{-iHt}$, where $\mc{O}$ is an initial operator and $H$ is the lattice Hamiltonian. This quantity is well defined for any quantum system and can be generalised to time evolving quantum states \cite{Balasubramanian:2022tpr,Kundu:2023hbk} as well as open quantum systems \cite{Liu:2022god,Bhattacharya:2022gbz,Bhattacharjee:2022lzy,Bhattacharya:2023zqt}. Since it captures dynamical information about the system under consideration, it has attracted a lot of attentions in the literature \cite{Fan:2022xaa,Fan:2022mdw,Caputa:2021sib,Jian:2020qpp,Rabinovici:2020ryf,Dymarsky:2021bjq,Bhattacharjee:2022ave,Bhattacharjee:2022vlt,Fan:2023ohh,Bhattacharyya:2023dhp,Afrasiar:2022efk,Pal:2023yik,Muck:2022xfc,Adhikari:2022whf,Adhikari:2022oxr,Avdoshkin:2022xuw,Camargo:2022rnt,
Hashimoto:2023swv,Iizuka:2023pov,Erdmenger:2023shk,Camargo:2023eev,Caputa:2024vrn,Bhattacharjee:2024yxj,Baggioli:2024wbz,Alishahiha:2024vbf}, or see the comprehensive review \cite{Nandy:2024htc}.

  In this paper, we would like to study the Krylov complexity in AdS/CFT correspondence. The operator growth in the boundary is holographically described by increase of the momentum of a free falling particle in the dual bulk geometry. In particular, in AdS$_2$ geometry, the operator size in SYK model is given by $s=P\tilde\beta$, where $P$ stands for the radial momentum of the infalling particle and $\tilde\beta$ denotes a local length scale, depending on the location of the particle. This simple relation has already established the close relation between the momentum of the infalling particle and the operator complexity (defined properly).

  More precisely, in Jackiw-Teitelboim (JT) gravity, the momentum generator is related to the length $L$ of the wormhole as
  \be P=\mc{J}\fft{dL}{dt} \,,\label{ads2}\ee
  where $\mc{J}$ is an energy scale in the theory. Previously, the wormhole length was proposed to give circuit complexity in the boundary \cite{Stanford:2014jda}. However, in a recent paper \cite{Rabinovici:2023yex}, by identifying the Hilbert space of JT gravity to that of the double scaled SYK model (near the ground state), it was shown that the bulk wormhole length in the semiclassical limit gives the boundary Krylov complexity in the continuum limit. This is however is an approximate dual since the continuum limit is a coarse graining description of quantum dynamics. Nevertheless, it predicts correct behaviors of K-complexity at both initial and late times.
  
 Inspired by \cite{Rabinovici:2023yex}, we would like to generalize the momentum-complexity relation (\ref{ads2}) to diverse dimensions. We propose
 \be \fft{dK}{dt}=P \,,\label{master}\ee
which relates the growth rate of K-complexity in the boundary to the bulk radial momentum of an infalling particle (the momentum is referred to that measured by a local stationary observer in the bulk). In this work, we will show that in AdS$_3$, the holographic K-complexity is perfectly matched with that of CFT$_2$. This is the first exact example of momentum-complexity correspondence. In diverse dimensions, our proposal  predicts correct behaviors of the K-complexity at both initial and late times in general AdS black hole background.

 The paper is organized as follows. In section 2, we briefly review the recursion method and establish universal features of K-complexity. In section 3, we try to generalize the relation (\ref{master}) to higher dimensions by taking the classical limit properly, without knowing the bulk dual of the K-complexity operator. In section 4, we adopt (\ref{master}) to calculate holographic K-complexity for explicit examples as well as general black hole background. In section 5, we briefly discuss the relation between our holographic K-complexity and the holographic circuit complexity. We briefly conclude in section 6.

\section{Universal behaviors of Krylov complexity}

To search holographic duals of Krylov complexity, let us first review the recursion method and establish the universal features of K-complexity.

Consider a lattice system described by the Hamiltonian $H$. The operator $\mc{O}$ evolves in time according to
\be \mc{O}(t)=e^{iHt}\mc{O} e^{-iHt}=\sum_{n=0}\fft{(it)^n}{n!}\tilde{\mc{O}}_n\,, \ee
where $\tilde{\mc{O}}_n$ stands for the nested commutators: $\tilde{\mc{O}}_1=[H,\mc{O}_0]\,,\tilde{\mc{O}}_2=[H,\tilde{\mc{O}}_1]\,,\cdots\,,\tilde{\mc{O}}_n=[H,\tilde{\mc{O}}_{n-1}]\,.$
However, the operator growth can also be viewed as the Schr\"{o}dinger evolution of a wave function, under the {\it Liouvillian} $\mc{L}\equiv [H\,,\cdot]$. One has
\be\label{obasis} |\mc{O}(t))=e^{i\mc{L}t}|\mc{O}_0 )= \sum_{n=0}\fft{(it)^n}{n!}|\tilde{\mc{O}}_n )\,, \ee
where $|\tilde{\mc{O}}_n )=\mc{L}^n|\mc{O}_0 ).$ To proceed, one has to introduce an inner product in the Krylov space appropriately. Generally, the original basis $\{|\tilde{\mc{O}}_n )\}$ will not be orthogonal (under the choice of inner product) but one can construct an orthonormal basis $\{|\mc{O}_n )\}$ using the Gram-Schmidt scheme.  We write
\be |\mc{O}(t))=\sum_{n=0}i^n\varphi_n(t)|\mc{O})_n\,, \label{knbasis}\ee 
where
\be |\mc{O}_0)=|\mc{O})\qquad\mm{and}\qquad  (\mc{O}_m|\mc{O}_n )=\delta_{mn} \,.\ee
Here we have assumed that the initial operator wave function $|\mc{O} )$ is normalized. Using the Gram-Schmidt scheme, the first basis vector will be evaluated as $|\mc{O}_1):=b_1^{-1}(\mc{L}-a_0)|\mc{O}_0)$, where $a_0=( \mc{O}_0|\mc{L}|\mc{O}_0)$ and $b_1=( \mc{O}_0|(\mc{L}-a_0)^2|\mc{O}_0 )^{1/2}$. At the $n-$th step, one has inductively
\bea\label{kbasis}
&&a_n=( \mc{O}_n|\mc{L}|\mc{O}_n )\,,\nn\\
&&|A_n):=(\mc{L}-a_{n-1})|\mc{O}_{n-1})-b_{n-1}|\mc{O}_{n-2})\,,\nn\\
&&|\mc{O}_n ):=b_n^{-1}|A_n )\,,\quad b_n:=( A_n|A_n)^{1/2}\,.
\eea
If $b_n=0$, the recursion stops, giving rise to a finite dimensional Krylov space. If not, one proceeds to the next step. The output of this procedure is not only an orthonormal basis $\{|\mc{O}_n )\}$ but also two sequence of positive numbers $\{a_n\,,b_n\}$, referred to as {\it Lanczos coefficients}. These coefficients have units of energy and can be used to measure time in the Heisenberg evolution. However, they play different roles, especially at initial and late times, despite that they both capture part of physical information of the system. Roughly speaking, $a_n$ acts as a phase factor in the wave function $\varphi_n$ whereas $b_n$ determines the amplitude, see our appendix \ref{appa} for more details.

According to (\ref{knbasis}), $\varphi_n(t)=i^{-n} (\mc{O}_n|\mc{O}(t))$ stands for a discrete set of (complex) wave functions and $p_n\equiv |\varphi_n|^2$ can be interpreted as probabilities, with the normalization
\be \sum_{n=0}^{\infty}|\varphi_n(t)|^2=1 \,.\ee
The Schr{\"{o}}dinger equation gives rise to a discrete set of equations
\be \pa_t\varphi_n=-ia_n\varphi_n+b_n\varphi_{n-1}-b_{n+1}\varphi_{n+1} \,,\label{phi0}\ee
subject to the boundary condition $\varphi_n(0)=\delta_{n0}$ and $a_{-1}=b_0=0=\phi_{-1}(t)$ by convention. This uniquely determines the wave functions $\varphi_n(t)$ for given Lanczos coefficients. 

The Krylov complexity is defined as
\be  K(t)=( \mc{O}(t)|\mc{K}| \mc{O}(t) )\,,\quad \mc{K}=\sum_{n=0}n |\mc{O}_n) ( \mc{O}_n| \,.
\ee
At initial times $t=0$, one finds to leading order
\be \varphi_0(t)=e^{ia_0t}+\cdots\,,\quad \varphi_1(t)=-e^{ia_0 t}b_1t+\cdots\,,\quad \varphi_2(t)=\fft12 e^{ia_0t}b_1b_2t^2+\cdots \,,\ee
Note the wave functions acquire a global phase factor associated to $a_0$. Evaluation of the Krylov complexity yields
\be K=b_1^2 t^2+\cdots\,,\ee
where dots stands for higher even power terms of $t$. The complexity grows quadratically to leading order, irrespective of the Lanczos coefficient $\{a_n\}$.

On the other hand, at late times, continuum limit analysis implies that the behavior of K-complexity heavily depends on the asymptotic growth of the Lanczos coefficient $\{b_n\}$, see our appendix \ref{appa} for details. In particular, for asymptotically linear growth of the Lanczos coefficient $b_n\sim \alpha n$, the complexity will grow exponentially $K(t)\sim e^{2\alpha t}$. 

Last but not least, since $\mc{O}(-t)=\mc{O}^\dag(t)$, the K-complexity turns out to be an even function of time $K(-t)=K(t)$, preserving the time reversal symmetry in the evolution. This will constrain validity of our momentum-Krylov complexity proposal in higher dimensions.

\section{The proposal for holographic Krylov complexity}

We have seen from (\ref{master}) that in AdS$_2$ geometry, the growth rate of Krylov complexity in the boundary is given by the momentum of the bulk infalling particle approximately. To generalize it to higher dimensions, we would like to study it from general aspects and show why it is possible.

Since operator growth is dual to a bulk infalling particle, the object we search that is dual to the K-complexity should be defined in the semiclassical limit. So let us first consider the classical counterpart of the K-complexity. Inspired by \cite{Rabinovici:2023yex}, we may introduce a classical complexity in terms of the distance traveled by the particle as
\be K_{cl}\equiv m L\,,\label{classicalcp}\ee
where $L$ is the (absolute) displacement and $m$ is the mass of the particle. This means the classical complexity measures the distance of the particle in units of $1/m$. Besides, we demand $K_{cl}$ grows quadratically at initial times as the quantum counterpart. This will be the case if the initial velocity is vanishing, consistent with our setup in holography. From (\ref{classicalcp}), we deduce
\be \fft{dK_{cl}}{dt}=P \,,\label{proposal}\ee
where $P$ stands for the momenta of the particle. To extend the result to a bulk infalling particle, we identify $P$ to the radial momentum of the particle, measured by a bulk stationary observer, according to holographic considerations. Notice that in this derivation we actually start with the semiclassical limit of a (properly defined) bulk length operator, which is however unknown in higher dimensional AdS geometries.

We can also discuss this by taking the classical limit of the growth rate of Krylov complexity directly. Consider a generic quantum state for a lattice system with Hamiltonian $H$. The growth rate of K-complexity can be evaluated as \cite{Fan:2022mdw,Hornedal:2022pkc}
\be \fft{dK}{dt}=-i\big\langle [H\,,\hat{n}]\big\rangle \,,\label{khh}\ee
where $\hat n\equiv \sum_{n=0} n\,|\mc{O}_n)(\mc{O}_n|$ denotes the K-complexity operator. The same relation holds for operator growth, with $H$ replaced by the Liouvillian $\mc{L}$. In the classical limit, the quantum average on the r.h.s will be replaced by the Possion bracket $\big\langle \big\rangle\rightarrow i \{\}_{\mm{PB}}$. Then if $\hat{n}$ is dual to a bulk length operator, the growth rate of K-complexity in the semiclassical limit will be given by the conjugate momenta\footnote{Here comes a subtlety: the momentum generator conjugate to the length operator is generally not given by $[H\,,\hat{n}]$  but the outcome of the expectation value $\big\langle [H\,,\hat{n}]\big\rangle$ equals to the momentum in the semiclassical limit. This is the case established in AdS$_2$ geometry.  }. This is how the relation (\ref{master}) is found in AdS$_2$ geometry.

 In higher dimensions, without a precise match between the Hilbert spaces of the boundary theory and that of the bulk theory, the quantity on the r.h.s of (\ref{khh}) remains unknown. Again, by holographic considerations, we propose that it is given by the radial momentum of the bulk infalling particle, leading to our proposal (\ref{master}). We will show that the proposal (\ref{master}) indeed predicts correct behaviors of K-complexity at both initial and late times at least. Here before moving to details, we briefly summarize the results: at initial times the holographic K-complexity grows quadratically as
 \be K=\fft{Et^2}{2\ell_{AdS}^2}+\cdots\,, \ee
whereas at late times it increases exponentially $K\sim e^{\fft{2\pi}{\beta}\,t}$ in general black hole background. These results are matched with universal behaviors of K-complexity very well.

\section{Testing the proposal}
To compute holographic K-complexity, we consider a free falling massless particle in a generally static and spherically symmetric black hole background. We work directly with the ingoing coordinate. One has in general $D$ dimensions
 \be
ds^2=-h(r)dv^2+2\sqrt{\ft{h(r)}{f(r)}}\,dvdr+r^2dx^idx^i\,,
\ee
where $i=2\,,3\,,\cdots\,,D-1$, $v=t+r_*(r)$ and $r_*=-\int_r^\infty \fft{\mm{d}r}{\sqrt{hf}}$ denotes the tortoise coordinate. The trajectory of the infalling particle is described by $v=\mm{cons}$. Without loss of generality, we set $v=0$, corresponding to the initial time $t=0$. Then the trajectory of the particle is determined by (a detail derivation is presented in our appendix \ref{appb})
\be\label{motion} t+r_{*}(r)=0 \,,\ee
which will in turn determine the growth of K-complexity. It turns out that the radial momentum of the particle is related to the energy $E$ as
\be P=\fft{E}{\sqrt{h(r)}} \,,\label{momentum1}\ee
where $E$ stands for the (proper) energy of the particle, which is conserved along the geodesic whereas $P$ refers to the momentum locally measured by a bulk stationary observer. Notice that we have taken absolute value of the particle momentum. Here it is worth emphasizing that the energy $E$ is not identical to the CFT energy $E_0$ in the boundary. According to the spirit of holography, we identify the CFT energy $E_0$ to the particle energy measured at the asymptotic AdS boundary. This gives
\be\label{ecft} E_0=\lim_{r\rightarrow \infty}\fft{E}{\sqrt{h(r)}}=\epsilon\,E\ell^{-1}_{AdS}\,, \ee
where the parameter $\epsilon$ stands for an ultraviolet cutoff, defined as $\epsilon=\lim_{r\rightarrow \infty}\ell_{AdS}^2/r$. While we will adopt the proper energy $E$ throughout this paper, the relation (\ref{ecft}) will help us to compare our holographic K-complexity with that in the boundary CFT.

Using the conjecture (\ref{master}), we deduce
\be K=\int_r^\infty dr'\,\fft{E}{\sqrt{f(r')}\,h(r')} \,,\label{geocp}\ee
where we have adopted the geodesic equation (\ref{motion}). Using this formula, we can compute the K-complexity straightforwardly for a given black hole background.

\subsection{AdS vacuum}
The simplest example is the AdS vacuum (in diverse dimensions), which has $h=f=r^2\ell_{AdS}^{-2}$. Solving the equation of motion (\ref{motion}) yields
\be \fft{\ell_{AdS}}{r}=\fft{t}{\ell_{AdS}} \,,\ee
which leads to
\be
P=\fft{E\ell_{AdS}}{ r}=\fft{Et}{\ell_{AdS}} \,.
\ee
From (\ref{master}), we deduce
\be
K=\fft{Et^2}{2\ell_{AdS}}\,,\label{initial}
\ee
where we always choose $K(0)=0$. It is intriguing that in this case the K-complexity grows quadratically in the full evolution. This is matched with the K-complexity in CFT$_2$ at the infinite temperature limit (see the next subsection). However, in general dimensions, it remains to be tested further. Nevertheless, the result predicts correct behavior of the K-complexity at initial times for holographic theories.

\subsection{BTZ black hole}
The next example is the BTZ black holes $h=f=(r^2-r_h^2)/\ell_{AdS}^2$, where $r_h$ stands for the horizon radius, which is related to the temperature $T=\beta^{-1}$ of the black hole as $r_h/\ell_{AdS}^2=2\pi/\beta$. Solving the equation of motion (\ref{motion}) yields
\be
r=r_h\coth{\big(\ft{2\pi}{\beta} t\big)} \,.
\ee
According to (\ref{momentum1}), the radial momentum of the particle is given by
\be
P=\fft{E\ell_{AdS}}{ r_h}\,\sinh{\big(\ft{2\pi}{\beta} t\big)}  \,.
\ee
Then our conjecture (\ref{master}) implies
\be K=\fft{2E}{\ell_{AdS}}\Big(\fft{\beta}{2\pi}\Big)^2\sinh^2{\big(\ft{\pi }{\beta}\,t \big)} \,.\label{btz}\ee
Interestingly, the result is perfectly matched with the K-complexity of CFT$_2$ at finite temperature \cite{Balasubramanian:2022tpr,Caputa:2024sux}, if one identifies the CFT energy $E_0$ to the particle energy $E$ according to the relation (\ref{ecft}). This gives the first exact example of momentum-complexity correspondence. 

\subsection{Higher dimensional black holes}

We move to study the K-complexity in higher dimensional black holes. Consider the Schwarzschild black holes, which have
\be h(r)=f(r)=\fft{r^2}{\ell_{AdS}^2}\Big(1-\fft{r_h^d}{r^d} \Big) \,,\ee
where $d=D-1$. The temperature is given by $ T=dr_h/4\pi \ell_{AdS}^2 $. Using (\ref{geocp}), we deduce
\be K=\fft{E \ell_{AdS}^3}{r_h^2}\int_0^z \mm{d}z'\,\fft{z'}{\big(1- z^{'d} \big)^{3/2}} \,,\ee
where $z\equiv r_h/r$. It will be convenient for us to work with a modular complexity
\be \widetilde{K}\equiv \fft{r_h^2}{E\ell_{AdS}^3}\,K \,.\ee
For later purpose, we will focus on the $D=5$ and $D=4$ dimensions.

\subsubsection{$D=5$ dimension}

\begin{figure}
  \centering
  \includegraphics[width=250pt]{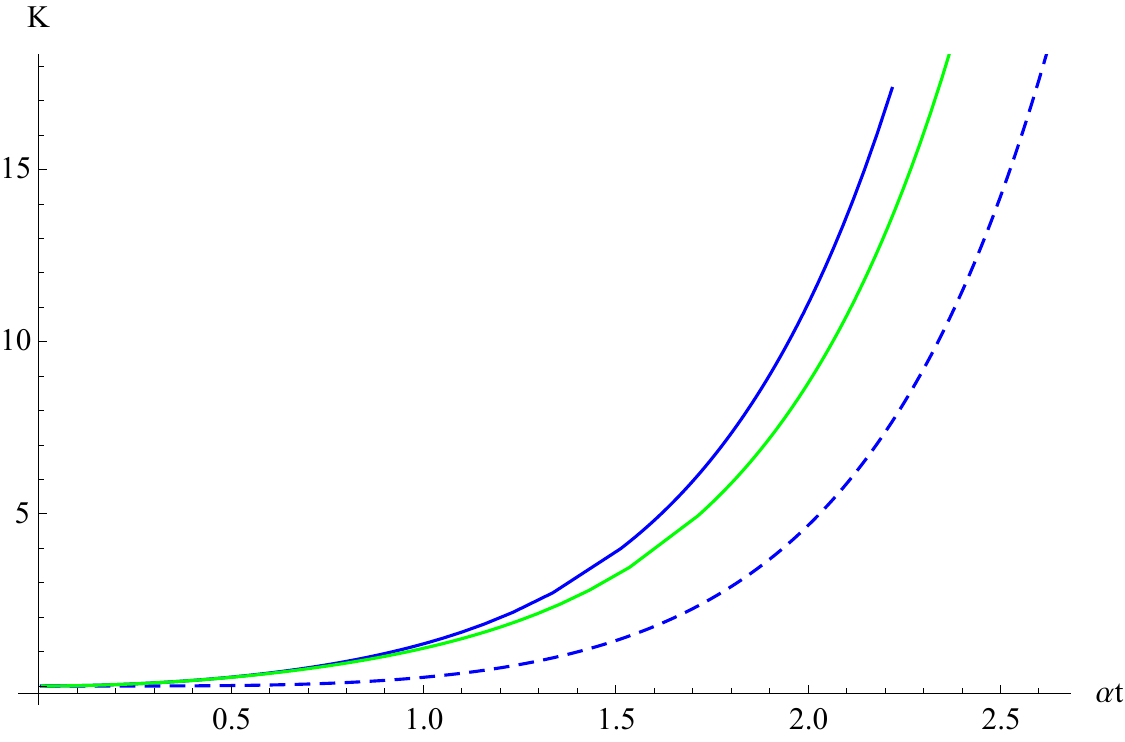}
  \caption{The Krylov complexity for five (green) and four (blue) dimensional Schwarzschild black holes. The dashed line is the refined complexity in the four dimension, which equals to half of the K-complexity at late times approximately. We have set $T=1/\pi$ and $E=2\ell_{AdS}$.}
  \label{schcp}
\end{figure}

In the five dimension, the tortoise coordinate can be solved as
\be r^*=-\fft{1}{4\pi T}
\Big(\pi-2\tan^{-1}\big(z^{-1}\big)+\log{\big(\ft{1+z}{1-z} \big)} \Big) \,,\ee
which then determines the motion of the infalling particle via (\ref{radialmotion}). The K-complexity can be evaluated as
\be \widetilde{K}=\fft{z(t)^2}{2\sqrt{1-z(t)^4}} \,.\label{ksch5}\ee
Numerical results show that the K-complexity soon grows exponentially, as seen in Fig. \ref{schcp}. On the other hand, we deduce at initial times
\be
K=\fft{E}{2\ell_{AdS}}\,t^2+\fft{\alpha^4 E}{20\ell_{AdS}}\,t^6-\fft{13\alpha^8 E}{3600\ell_{AdS}}\,t^{10}+\fft{31\alpha^{12} E}{31200\ell_{AdS}}\,t^{14} +\cdots \,,
\ee
where $\alpha=\pi T$. Note that all odd powers of $t$ is absent. In fact, according to (\ref{ksch5}),  the K-complexity preserves the time reversal symmetry in the full evolution because of $z(-t)=-z(t)$. 

\subsubsection{$D=4$ dimension}

For the $D=4$ dimension, one finds
\be r^*=-\fft{\sqrt{3}}{4\pi T}\Big(\fft{\pi}{2}-\tan^{-1}\big(\ft{2+z}{\sqrt{3}z} \big) \Big)
-\fft{1}{8\pi T}\log{\Big( \ft{1+z+z^2}{1-2z+z^2} \Big)} \,.\ee
The K-complexity can be solved in terms of hypergeometric functions
\be \widetilde K=\fft{z^2(t)}{6}\Big( \fft{4}{\sqrt{1-z^3(t)}}-{}_2F_1\big[\fft12\,,\fft23\,,\fft53\,,z^3(t) \big] \Big) \,. \ee
Again from numerical results we find the complexity soon grows exponentially, as shown in Fig. \ref{schcp}. However, a significant difference from the previous case is that $K$ does not preserve the time reversal symmetry any longer. One has at initial times
\bea
K&=&\fft{E}{2\ell_{AdS}}\,t^2+\fft{16\alpha^3 E}{135\ell_{AdS}}\,t^5-\fft{64\alpha^6 E}{5103\ell_{AdS}}\,t^{8}+\fft{65536\alpha^{9} E}{7577955\ell_{AdS}}\,t^{11}+\cdots \,.
\eea
The presence of odd powers implies that our proposal (\ref{master}) in this case does not produce the correct initial behavior of K-complexity beyond  the leading order.

In fact, there is no reason to expect that the time reversal symmetry of the K-complexity is respected by our proposal in generally asymptotically AdS spacetimes. Previously we have seen that to arrive at (\ref{master}), we have to take the semiclassical limit which is somewhat subtle. As a consequence, even in AdS$_2$ geometry, it only predicts the leading order behavior of the K-complexity at initial and late times whereas it generally loses predictive power at intermediate times. In view of this, it is particularly interesting that in AdS$_3$, our proposal (\ref{master}) matches with the boundary theory perfectly.

To remedy this issue partly, we may introduce a refined momentum
\be P_{\mm{re}}(t)\equiv \fft12\Big( P(t)-P(-t) \Big) \,,\label{remomentum}\ee 
and the proposal (\ref{master}) is promoted to be 
\be \fft{dK_{\mm{re}}}{dt}=P_{\mm{re}} \,,\label{promotemaster}\ee
where $K_{\mm{re}}$ stands for the refined K-complexity. In cases where the time reversal symmetry of the complexity is preserved, $P(-t)=-P(t)$ and hence the refined momentum (and complexity) simply reduces to the original momentum (and complexity) of the infalling particle. Generally since one always has $P_{\mm{re}}(-t)=-P_{\mm{re}}(t)$, the time reversal symmetry will be always preserved by the refined momentum $K_{\mm{re}}(-t)=K_{\mm{re}}(t)$. However, there are some shortcomings for this notion. The first is the leading order behavior of the complexity at late times will generally be changed by a proportionality factor, see the numerical results of four dimensions in Fig. \ref{schcp}. The second is it predicts wrong behavior of the complexity in the post scrambling regime, see the next subsection.

\subsection{Universal results at late times }

Let us study the behavior of holographic K-complexity at late times for general black hole background. In the near horizon region, the particle falls as
\be r-r_h\propto e^{-\fft{4\pi}{\beta}\,t} \,.\ee
Substituting the result into (\ref{momentum1}), one finds
\be P=P_*\, e^{\fft{2\pi}{\beta}\,t} \,,\ee
where $P_*$ is a constant. Then we obtain
\be K=\fft{\beta}{2\pi}\,P\propto e^{\fft{2\pi}{\beta}\,t} \,.\label{late}\ee
The result matches with the exponential growth of K-complexity for fast scramblers at long times. Here notice that the exponent $\alpha=\pi/\beta$ is universal to holographic theories at finite temperatures. These results are valid to the refined momentum and complexity as well except for that $P_{\mm{re}}\rightarrow c\, P\,,K_{\mm{re}}\rightarrow c\, K$ in the near Rindler regime, where $c\rightarrow 1/2$ when $K$ does not respect the time reversal symmetry.  

To check the above result further, we would like to see whether the Lanczos coefficient $b_n$ grows asymptotically linearly. To address this issue, we adopt the continuum limit at late times. In this approximation, the wave function $\varphi_n$ moves ballistically with a velocity $v=2b_n$ and the K-complexity simply measures the location of the peak $K\sim n$. One has
\be \fft{dK}{dt}\Big|_{t\rightarrow \infty}=2b(K) \,.\label{continuum}\ee
Combing the result with our conjecture (\ref{master}) and the result (\ref{late}), one finds at late times
\be b(K)=P/2=\alpha K \,.\ee
Indeed, the Lanczos coefficient grows asymptotically linearly, with the correct rate $\alpha=\pi/\beta$ for fast scramblers dual to general AdS black hole background.

Up to now, we have not considered the back-reaction effects of the particle in the near Rindler regime. This generally corresponds to the post scrambling regime in which the Lanczos coefficient $b_n$ approaches to a constant $b_n\rightarrow b$. In this case, the K-complexity will increase linearly $K\propto bt$ \cite{Barbon:2019wsy}. On the other hand, beyond the scrambling time, the particle falls into the black hole interior and the radial momentum $P$ will approach to a constant \cite{Lin:2019qwu}. Therefore, our holographic K-complexity will increase linearly as well. Again it matches with the universal behavior of K-complexity very well. In contrast,  the refined momentum introduced in (\ref{remomentum}) will vanish in the black hole interior, leading to a constant $K_{\mm{re}}$ in the post scrambling regime. This raises the question: whether a proper refined K-complexity obeying both the universal behaviors of the complexity and the time reversal symmetry exists.

To end this section, we conclude that at late times
\be\label{twolate}
K(t)\sim \left\{\begin{array}{ll}
e^{\fft{2\pi}{\beta}t}\,,\qquad\qquad \qquad\mathrm{for}\quad  t_d< t < t_*\,,  \\
b\, (t-t_*)\,,\qquad \qquad\mathrm{for}\quad  t>t_*\,,     \\
\end{array}
\right.
\ee
where $t_d\,(t_*)$ stands for the dissipative (scrambling) time, respectively. It is interesting that the results are qualitatively the same as those of circuit complexity.

\section{The relation to circuit complexity}

According to the Complexity-Volume (CV) duality \cite{Stanford:2014jda}, the wormhole volume of AdS black holes is dual to the circuit complexity $\mc{C}$ in the boundary theories. Recall Eq.(\ref{ads2}), this implies in AdS$_2$ geometry
\be \fft{d\mc{C}}{dt}=2\pi P \,.\label{cc1}\ee
By quantum circuit consideration, the result was previously generalised to higher dimensions in \cite{Susskind:2020gnl}
\be \fft{\lambda}{2\pi}\fft{d\mc{C}}{dt}=2\pi P \,,\label{cc2}\ee
where $\lambda$ stands for the dimensionless wavelength of excitations in the boundary, which is dual to bulk quantities as $\lambda=\tilde\beta/\ell_{AdS}$, where $\tilde\beta$ is a local length scale, depending on the location of the particle (in the near Rindler regime $\tilde\beta=\beta$). This is referred to as momentum-complexity correspondence in the literature. Comparing the above results with our proposal (\ref{master}) for the K-complexity, we find a general connection between the two quantities
\be \fft{\lambda}{4\pi^2}\fft{d\mc{C}}{dt}=\fft{dK}{dt} \,.\label{connection}\ee
In particular, in AdS$_2$ geometry $\Delta\mc{C}(t)=2\pi K(t)$, meaning the increase of circuit complexity is equal to that of the K-complexity, up to $2\pi$ times. In higher dimensions, their relation will be implicitly determined by (\ref{connection}). Here we are interested in comparing their universal behaviors in certain time regimes.
One has at initial times $ \Delta \mc{C}(t)=2\pi E t$, whereas the K-complexity grows quadratically as (\ref{initial}). The two increases with different initial rates. However, at late times they both grow exponentially in the scrambling regime and increase linearly in the post scrambling regime \cite{Susskind:2020gnl}. In view of this, we may conclude that the two can equivalently characterize the dynamical evolution of fast scramblers at long times.
This also provides a possible candidate for the CFT dual of holographic circuit complexity as well as the gravity duals of K-complexity, see recent developments in momentum complexity correspondence \cite{Barbon:2019tuq,Barbon:2020olv,Barbon:2020uux}.

\section{Conclusion}

In this paper, we relate the growth rate of Krylov complexity to the radial momentum of an infalling particle in AdS geometry. By studying the proposal in general black hole background, we show that it well captures the universal behaviors of K-complexity at both initial and late times. This gives us confidence that the proposal can be considered as an approximate dual of the Krylov complexity at least. In particular, in AdS$_3$, our holographic K-complexity exactly matches with the K-complexity of CFT$_2$, giving the first exact example of momentum-complexity correspondence. Application of the proposal to more examples and clarifying its deep relation to holographic circuit complexity deserve further investigations. Following recent developement \cite{Ageev:2018msv}, it will be also interesting to generalize the proposal to the growth of operator with a global charge, which is described by charged particle falling to the Reissner-Nordstrom black hole. We leave these to future search.\\

\section*{Acknowledgments}

Z.Y. Fan was supported in part by the National Natural Science Foundations of China with Grant No. 11805041 and No. 11873025.\\
~~~\\
~~~\\
\textbf{{\textcolor{blue} {{\it Note added}. When our draft is in preparation, the paper 2410.23334  is published on arXiv preprint. In that paper, the momentum-Krylov complexity correspondence is (only) examined in AdS$_3$ using an infalling massive (rather than masless) particle. However, we both reproduce the Krylov complexity of the dual CFT$_2$ exactly by proper identification for the CFT energy. }}}

\appendix
\section{Continuum limit analysis at long times}\label{appa}

Continuum limit analysis is good at capturing the leading long time behaviors of Krylov complexity for irreversible process using coarse grained wave functions. To solve the Schr\"{o}dinger equation in this approach
\be \pa_t\varphi_n=-ia_n\varphi_n+b_n\varphi_{n-1}-b_{n+1}\varphi_{n+1} \,,\label{phi}\ee
we define a continuous coordinate $x=\epsilon n$ as well as a velocity $v(x)=2\epsilon b_n$, where $\epsilon$ is a lattice cutoff. The coarse grained wave function is simply defined as $\varphi(x\,,t)=\varphi_n(t)$. We take $a(x)=a_n$. The wave equation (\ref{phi}) gives
\be \partial_t\varphi(x\,,t)=-ia(x)\varphi(x\,,t)+\fft{1}{2\epsilon}\Big[ v(x)\varphi(x-\epsilon)-v(x+\epsilon)\varphi(x+\epsilon) \Big] \,. \ee
Expanding the equation in powers of $\epsilon$ yields to leading order
\be\label{expansion} \partial_t\varphi=-ia(x)\varphi-v(x)\partial_x\varphi-\fft12 \partial_x v(x)\varphi+O(\epsilon) \,.\ee
This is a chiral wave equation with a position-dependent velocity $v(x)$ and a complex mass $m(x)=\fft12 \partial_x v(x)-ia(x)$. This will be clear if we work with a new frame $y$, defined as $y=\int dx/v(x)$ and a rescaled wave function
\be \tilde\phi(y\,,t)=\sqrt{v(y)}\,\varphi(y\,,t)\,\mm{exp}\big[ -i\int^y a(y')dy' \big] \,.\ee
The wave equation simplifies to
\be\label{chiralwave} (\partial_t+\partial_y)\tilde\phi(y\,,t)=0+O(\epsilon) \,.\ee
The general solution is simply given by
\be \tilde\phi(y\,,t)=\tilde\phi_i(y-t) \,,\ee
where $\tilde\phi_i(y)=\tilde\phi(y\,,0)$ stands for the initial amplitude. The result implies that to leading order the wave function moves ballistically at long times. Notice that the Lanczos coefficient $a_n$ (for large $n$) simply acts as a constant phase in the original wave function. This might be oversimplified but is enough to estimate the growth of K-complexity at long times.

To derive Krylov complexity, we first rewrite the normalization condition
\be 1=\sum_{n}|\varphi_n(t)|^2=\fft{1}{\epsilon}\int \mm{d}x\, |\varphi(x\,,t)|^2=\fft{1}{\epsilon}\int \mm{d}y\, |\tilde\phi_i(y)|^2  \,.\ee
Evaluation of the complexity yields
\be\label{cksk}
K(t)=\sum_{n}n |\varphi_n(t)|^2=\fft{1}{\epsilon}\int\mm{d}y\, \fft{x(y+t)}{\epsilon}\,\, 
|\tilde\phi_i(y)|^2 \,.
\ee
Once the transformation between the two frames is known, we are able to extract the leading time dependence of the quantity immediately. One has at long times $K\sim x(t)/\epsilon$. Typically consider the asymptotically linear growth of the Lanczos coefficient $b_n\rightarrow \alpha n$. The Krylov complexity will grow exponentially at late times $K(t)\sim e^{2\alpha t}$. On the other hand, if the Lanczos coefficient approaches to a constant $b_n\rightarrow b$, the complexity will grow linearly $K(t)\sim 2 bt$.

\section{Free falling particles in AdS spacetimes}\label{appb}
Considering the generally static and spherically symmetric metric in $D$ dimensions
\be ds^2=-h(r)dt^2+dr^2/f(r)+r^2 dx^idx^i\,,\ee
where the metric functions $h(r)\,,f(r)$ have not been specified. In such a curved spacetime, the motion of a massless particle is characterized by a null geodesic. We only consider a free falling particle along the radial direction. The four momentum of particle in the canonical coordinates is specified by $\mc{P}_\mu=(-E\,,\mc{P}_r\,,0)$, where $E$ and $\mc{P}$ is the energy and the radial momentum of the particle. One has the normalization
\be 0=\mc{P}^\mu \mc{P}_\mu=-|g^{tt}|E^2+g^{rr}\mc{P}_r^2\,. \label{mmnormal}\ee
On the other hand, the local four momentum is the same quantity measured by a stationary observer in the bulk and obeys the same normalization. We denote it by $P_\mu=(-\mc{E}\,,P_r\,,0)$. One has
\be \mc{E}=\sqrt{|g^{tt}|} E\,,\quad P_r=\sqrt{g^{rr}}\mc{P}_r\,.\label{localmomentum}\ee
Notice that $E$ is conserved along the geodesic whereas the locally measured energy $\mc{E}$ is position (or time) dependent. However, the later gives the CFT energy $E_0$ at the asymptotic AdS boundary
\be E_0=\epsilon\, E\,\ell^{-1}_{AdS} \,,\ee
where the regularization parameter is defined as $\epsilon=\lim_{r\rightarrow \infty}\ell^2_{AdS}/r$.

The four velocity of the particle is given by
\be U^\mu=\fft{dx^\mu}{d\tau}=(\dot{t}\,,\dot{r}\,,0) \,,\ee
where $\tau$ is the proper time and a dot denotes the derivatives with respect to $\tau$. By definition, one has
\be \mc{P}_\mu=U_\mu=g_{\mu\nu}U^\nu \,,\ee
such that
\be E=h(r)\dot t\,,\quad \mc{P}_r=\fft{\dot r}{f(r)}\,.\label{mmtinf}\ee
The Lagrangian is given by
\be \mathcal{L}(x\,,\dot x)=\fft 12 g_{\mu\nu}\dot{x}^\mu \dot{x}^\nu=\fft12\Big(-h(r)\dot{t}^2+\fft{\dot{r}^2}{f(r)} \Big) \,,\ee
Since the Lagrangian does not explicitly depend on $t$, there is a first integral
\be E=-\fft{\partial \mathcal{L}}{\partial \dot t}=h(r)\dot t\,,\ee
which is a conserved quantity along the null geodesic. According to the normalization (\ref{mmnormal}), one finds
\be \mc{P}_r=\fft{\dot r}{f(r)}=-\fft{E}{\sqrt{h(r)f(r)}} \,,\ee
where a minus sign is introduced on the r.h.s because $r$ decreases as the particle moves towards the center. This leads to
\be \dot r=-\sqrt{\fft{f(r)}{h(r)}}\,E \,,\ee
as well as the local radial momentum
\be P_r=\fft{\dot r}{\sqrt{f(r)}}=-\fft{E}{\sqrt{h(r)}}\,.\label{localmomentum2}\ee
Combining the above results, we deduce the motion of the particle in real time
\be \fft{dr}{dt}=-\sqrt{h(r)f(r)} \,. \label{motionreal}\ee
It follows that
\be t=-r^*(r)=\int_{r}^{\infty}\fft{dr}{\sqrt{h(r)f(r)}} \,,\label{radialmotion}\ee
where $r^*$ is the tortoise coordinate and we place the particle at cut-off surface $\infty$ at initial time $t=0$. Finally, it should be emphasized that integrating the momentum $P_r$ over the proper time $\tau$ (rather than the real time) yields the proper length $L_p$ travelled by the particle $L_p=\int d\tau\,P$, where $P=|P_r|$.

\end{document}